\documentclass[aps,prl,twocolumn,showpacs,amsmath,amssymb]{revtex4}

\usepackage{dcolumn}
\usepackage{bm}

\begin{document}

\title{Disorder resistivity of solid neutron-star matter}

\author{P. B. Jones}
\

 \email{p.jones1@physics.ox.ac.uk}
\affiliation{Department of Physics, University of Oxford\\
Denys Wilkinson Building, Keble Road, Oxford, OX1 3RH, England.}%

\date{\today}

\begin{abstract}
Lower limits are found for the disorder electrical resistivity of solid
neutron-star
matter in the neutron-drip region which is amorphous and heterogeneous in
nuclear charge.  This temperature-independent resistivity, large compared with
that produced by phonon scattering, has direct consequences for theories of
neutron-star magnetic field generation and evolution.  
\end{abstract}

\maketitle

Following the discovery of neutron stars in 1967, very many papers have been
published containing theories of the nature and origin of their
magnetic flux distributions.  The
original view was that the distributions are fossil fields amplified during
gravitational collapse.  A recent variant of this idea is field
generation by rapid dynamo action during a brief interval of convective motion
immediately after that process \cite{td93}.  But some authors have doubted that
amplified fossil fields can be of the order of magnitude inferred from
measured pulsar periods and their time-derivatives, and have considered
thermoelectric field generation in the solid crust and liquid surface region
of the star \cite{uy80, bah83, wg96} during the early years after its
formation.  This Letter shows that disorder electrical resistivity is
present in solid
neutron-star matter.  It is temperature-independent and so large that, with
the impurity resistivity, it
severely constrains other than fossil-field theories.
 
At times when thermoelectric processes have
become negligible, and with no movement of matter, the
evolution of the magnetic flux density ${\bf B}$ in the solid crust
is given by the
Maxwell induction equation with Ohm's law and neglect of displacement
currents,
\begin{eqnarray}
\frac{\partial {\bf B}}{\partial t} & = & -\nabla\times\left(\frac{c^{2}}
{4\pi\sigma}\nabla \times{\bf B}\right) \nonumber\\
  &   & - \nabla\times
\left(\frac{c}{4\pi ne}\left(\nabla\times{\bf B}\right)\times{\bf B}\right),
\end{eqnarray}
where $\sigma$ is the zero-field conductivity and $n$ the electron number
density.  Eq. (1) assumes that $B$ is not so large in relation to $n$ that
classical conductivity fails (see, for example, Ref. \cite{p99}).
The existence of disorder resistivity leads to small values of $\sigma$,
renders negligible the term in (1) bilinear in ${\bf B}$ representing
Hall drift, and allows unambiguous calculation of field evolution within
the crust.

Previously, values of $\sigma$ have been derived from calculations of
electron-phonon
scattering in a homogeneous {\it bcc} lattice \cite{um92}, and from
scattering \cite{yu80, gyp01, ik93} by fractional concentrations $c_{i}$ of
impurity charges $Z_{i}$ with mean $\bar{Z}$ and mean square deviation
$Q = \sum_{i}c_{i}\left(Z_{i} - \bar{Z}\right)^{2}$.
The latter has been treated as a free parameter, usually assumed to
satisfy $Q\ll 1$.
For these $\sigma$, the Hall term in (1) 
is, in many cases, larger than the
ohmic dissipative linear term \cite{j88, gr92}.  Eq. (1) has some
similarity with the vorticity equation for a viscous incompressible fluid
(see, for example, Ref. \cite{b67}) and a number of authors \cite{gr92,
rg02,hr04} have investigated the existence
of a cascade to higher-wavenumber components
analogous with the Kolmogorov cascade in fluid
turbulence.  This would have a significant effect on evolution because,
although Hall drift is not in itself dissipative,
the ohmic dissipation rate for any flux-distribution component is dependent
on the square of its wavenumber.

It must be emphasized that the references
cited so far are not complete but provide an introduction
to a very large literature on a subject of some complexity.
Solutions of Eq. (1) are not simple if the Hall term is appreciable.
Also, if $\sigma$ is derived from electron-phonon resistivity, it is
so highly temperature-dependent that the cooling of the star and the
evolution of its magnetic flux distribution must, in principle,
be calculated in parallel.  The present work shows that disorder becomes the
dominant source of
resistivity as the interior cools below $10^{9}$ K and so obviates these
problems.

There is now evidence that the assumption of a {\it bcc}
lattice, on which calculations
of $\sigma$ have been based, is not correct.  Weak-interaction equilibrium
is not reached during the formation and cooling of the neutron star so
that, as a consequence of shell structure and of the small differences
in formation enthalpy between nuclei of different $Z$, the solid is both
heterogeneous and amorphous \cite{j01,j04}. (This work applies to
neutron stars that have not undergone significant mass-accretion.  The
high mass-transfer rates in certain binary systems can completely replace
the natural solid crust with accreted matter of heterogeneous $Z$
\cite{sbcw99}.)  It is well known that amorphous
systems have high and substantially temperature-independent
resistivities (see, for example, Ref. \cite{rao}).  In solid
neutron-star matter, disorder is the most important source of resistivity
and does not seem to have been considered previously.

If the weak-interaction strength were (hypothetically) adjusted so as to
allow the attainment of equilibrium, an amorphous and $Z$-heterogeneous
solid would transform to a {\it bcc} structure with homogeneous nuclear
charge and a population of lattice defects.  Thus an amorphous
structure with minimum disorder can be obtained by the reverse
transformation, starting from a defect-free {\it bcc} lattice
and sequentially changing each nuclear charge $\bar{Z}\rightarrow Z_{i}$
randomly so as to give fractional concentrations $c_{i}$.  This procedure
has been carried out for a system of 24357 {\it bcc} lattice sites contained
within a spherical volume using distributions $c_{i}$ of $Z_{i}$ found by the
methods of Ref. \cite{j01, j04}.  These are based on equilibrium {\it bcc}
lattices with charge $Z_{CLDM}$ whose parameters have been obtained
at temperature $T=0$ for each
of the matter densities in Table I using the
compressible liquid-drop model of Lattimer et al \cite{lprl85}. The densities
selected are in the neutron-drip region and represent an interval equivalent
to most of the depth of the crust.  Calculations of formation enthalpies for
nuclei with $20 \leq Z_{i} \leq 50$ in equilibrium with the neutron continuum
\cite{j01,j04} give values of the $c_{i}$ for thermal equilibrium,
at an initial temperature $T_{o}=5\times 10^{9}$ K, which is near
the melting
temperature $T_{m}$ of the solid.
There is movement away from weak-interaction equilibrium during the
early stages of cooling and the fractional concentrations $c_{i}$ evolve
to frozen-in values giving the parameters $Q$ of Table I.  This 
produces a small shift $\delta Z$ in nuclear charge so that
the mean charge of Table I is $\bar{Z} = Z_{CLDM} + \delta Z$.
Because there are considerable uncertainties in obtaining the shell-corrected
nuclear formation enthalpies, a range of possibilities is considered
\cite{j04}.  Rows labelled sp in Table I are based on values of the
$c_{i}$ found from formation enthalpies given by the Strutinski procedure
for shell-correction (see, for example, Ref. \cite{rs80}) with proton
pairing.  Rows labelled p are for $c_{i}$ given by formation enthalpies
with proton pairing but no shell correction.

Each change of nuclear charge
$\bar{Z}\rightarrow Z_{i}$ produces displacements of all other lattice
sites in the system considered.  Our assumption here is that these are
identical with those found in Ref. \cite{j01} for an isolated point-defect
in a homogeneous {\it bcc} lattice.  (This, of course, is not strictly
correct because as the sequence of transformations $\bar{Z}\rightarrow Z_{i}$
is made, the neighbours of a given nucleus become heterogeneous in charge
and their positions deviate from those of the {\it bcc} lattice.)  These
displacements were calculated for each of the neighbour sets $\alpha=1-3$,
but for $\alpha\geq 3$, they are assumed equal to those produced by a
point defect in an isotropic incompressible elastic medium.  The volume
change represented by these displacements is determined, principally, by
the properties of the Coulomb-electron stress tensor in the neutron-drip
region whose isotropic components are between one and two orders of
magnitude larger than the off-diagonal, and by the fact that
$\kappa a \sim 1$,
where $\kappa$ is the electron screening wavenumber \cite{j01,j04}.  Thus
the displacements and volume change are such that
the electron density, averaged over a volume of the order of
a Wigner-Seitz cell, adjusts to values almost exactly equal to those
of the undisturbed lattice with Fermi wavenumber $k_{Fe}$.
The result of calculating the total displacement
for each lattice site in the system considered is that
the radial density distribution of nuclear positions relative to the origin
site changes from a sum of $\delta$-functions
in the undisturbed {\it bcc} system to a sum of gaussian functions
of approximately constant variance $w^{2}$.  There is also a change in
the mean spatial density of sites, but it is statistically consistent
with zero.  Values of $w$ have been
calculated for each matter density and are given in Table I.

Our calculation of disorder resistivity follows closely the classic
paper of Ziman \cite{z61} on liquid metals.  Comparison
with the exact electron scattering matrix element \cite{yrw54}
as a function of the momentum transfer $q$ indicates that for the size of 
momentum transfer allowed in the neutron-drip region, $0 < q < 2k_{Fe}$,
the Born approximation is adequate
for scattering by the potential inside an
electrically neutral Wigner-Seitz cell. The Fourier transform of the
charge density is
$ZeF(q)$.  It assumes the uniform proton density of
a CLDM nucleus \cite{lprl85} and uniform electron density inside a
charge-neutral spherical
cell whose radius is defined by $\bar{Z}$ and $k_{Fe}$.
The scattering amplitude for the whole system is then a sum of two
terms; one dependent on $\bar{Z}$ and the other on the differences
$(Z_{i} - \bar{Z})$.  The static structure function derived from the
first term,
\begin{eqnarray}
S_{d}(q) = \frac{1}{N}\left|\sum_{s=1}^{N}
{\rm e}^{{\rm i}{\bf q}\cdot{\bf r}_{s}}\right|^{2},
\end{eqnarray}
which vanishes in the absence of disorder,
has been obtained directly, to terms of order $q^{2}w^{2}$, from the Fourier
transform of the radial density distribution of the nuclear positions
${\bf r}_{s}$
Nuclei of any specific charge $Z_{i}$ are randomly distributed
and the set of such nuclei has static structure function $S_{z} = 1$.

\begin{table}
\caption{Disordered systems at matter density $\rho$ have mean nuclear
charge $\bar{Z}$ and electron Fermi wavenumber $k_{Fe}$ giving an
equivalent {\it bcc} lattice constant $a$.  The parameter $Q$ is the
mean square nuclear charge deviation and
$w^{2}$ is the variance of the gaussian radial distributions of nuclear
positions relative to the origin site.
Rows labelled sp
are for the cases in which nuclear concentrations $c_{i}$ have
been obtained by the procedures of Ref. \cite{j04} with shell corrections
and proton pairing.  Rows labelled p are for $c_{i}$ given by proton
pairing alone.}
\begin{ruledtabular}
\begin{tabular}{l c c c c c r}
   & $\rho$  & $\bar{Z}$ & $k_{Fe}$ & $a$
 & Q & $w$  \\
  & ($10^{13}$g cm$^{-3}$) &   & ($10^{13}$ cm$^{-1}$) & ($10^{-13}$ cm) & 
   & ($10^{-2}a$) \\
\hline
 sp & 1.6 & 37.8 & 0.231 & 56.6 & 11.9 & 1.47  \\
 sp & 3.7 & 35.3 & 0.286 & 44.7 & 6.0 & 1.80 \\ 
 sp & 8.8 & 39.0 & 0.363 & 36.4 & 19.0 & 1.90 \\
 p & 1.6 & 34.6 & 0.231 & 55.0 & 5.0 & 1.40 \\
 p & 3.7 & 33.8 & 0.286 & 44.1 & 17.4 & 2.46 \\
 p & 8.8 & 34.4 & 0.363 & 34.9 & 24.3 & 3.19 \\
\end{tabular}
\end{ruledtabular}
\end{table}
There is no interference between the two amplitude terms and so the
resistivity $\mathcal{R}$ is
\begin{eqnarray}
\mathcal{R}=\mathcal{R}_{z}+\mathcal{R}_{d}= 
\frac{4\pi e^{2}}{\bar{Z}c\mu_{e}}\left(Q\Lambda_{z}+\bar{Z}^{2}
\Lambda_{d}\right),
\end{eqnarray}
where $\mu_{e}$ is the electron chemical potential.
The modified Coulomb integrals are,
\begin{eqnarray}
\Lambda_{d,z}=\int^{2k_{Fe}}_{0}\frac{dq}{q}
\left(1-\frac{q^{2}}{4k_{Fe}^{2}}\right)F^{2}(q)S_{d,z}(q){\rm e}^{-2W},
\end{eqnarray}
in which the final term, the Debye-Waller factor, can be set equal to unity
at all $q\leq 2k_{Fe}$ for the relevant $T\ll T_{D}$, where $T_{D}$ is
the Debye temperature of a
neutron-drip {\it bcc} lattice.  The first term in Eq. (3) is identical
with the standard expression for impurity resistivity but the Table II
values of $\mathcal{R}_{z}$ obtained from it are smaller than those of
Ref. \cite{gyp01, ik93} owing to the different assumption in the present
work concerning $F(q)$ .  (The assumption of \cite{gyp01}, that $F(q)$ is
the nuclear form factor divided by the small-$q$ limit of the longitudinal
static dielectric constant, gives the bracketed $\mathcal{R}_{z}$ values.)
The second is the disorder resistivity and, even for the minimum disorder
case, it is the larger term given the $F(q)$ of this Letter,
as shown in Table II.

It is difficult to be specific about examples of greater disorder than
the minimum defined here.
One possibility, which we arbitrarily define as maximum disorder,
would resemble the inital stages of component separation
in which small volumes of homogeneous lattice exist as irregular polyhedra,
with random orientations. It is assumed
that their linear dimensions are so small that the wave function of
an electron is almost independent of the reciprocal lattice vectors of
the polyhedron within which it is moving.  In this case, and under the
assumption of random orientation, the structure function (2) becomes
\begin{eqnarray}
S_{d}(q) = \frac{1}{N}\sum_{\alpha}N_{\alpha}S_{\alpha}(q),
\end{eqnarray}
where $S_{\alpha}$ is the structure function for a single polyhedron
of $N_{\alpha}$ nuclei.  Order of magnitude estimates of the resistivity
can be found by substituting the mean $\bar{S}_{\alpha}$ for $S_{d}$ in
Eq. (4).  Values of $S_{\alpha}$ have been calculated for a number
of trial irregular polyhedra containing $\sim 10^{3}$ nuclei.  Typical values 
give the resistivities $\mathcal{R}_{d}^{max}$
in the final column of Table II, which are not sensitive to polyhedron size.
Their order of magnitude shows that more disordered structures
can produce very high resistivity.  On this basis, we regard the values
of $\mathcal{R}$ in the penultimate column of Table II as lower limits,
though it has to be emphasized that they are subject to considerable
uncertainties arising from the shell-correction procedure, as is shown
by the differences between those labelled p and sp.

\begin{table}
\caption{Resistivities are listed for systems with the properties
given in Table I, $\mathcal{R}_{d}^{min}$ from the effect of minimum
disorder and $\mathcal{R}_{z}$ from charge heterogeneity;
$\mathcal{R}=\mathcal{R}_{d}^{min}+\mathcal{R}_{z}$.  The bracketed
values of $\mathcal{R}_{z}$ are for the form factor assumed in
\cite{gyp01}. The final
column gives an estimate of the resistivity arising from maximum
disorder as defined here.}
\begin{ruledtabular}
\begin{tabular}{l c c c c r}
  & $\rho$ & $\mathcal{R}_{d}^{min}$ & $\mathcal{R}_{z}$ &
  $\mathcal{R}$ & $\mathcal{R}_{d}^{max}$  \\
  & ($10^{13}$ g cm$^{-3}$) & ($10^{-24}$s) & ($10^{-24}$s) & ($10^{-24}$s)
  & ($10^{-24}$s)  \\
\hline
 sp & 1.6 & 0.49 & 0.31 (0.73) & 0.80 & 35  \\
 sp & 3.7 & 0.39 & 0.11 (0.29) & 0.50 & 21  \\
 sp & 8.8 & 0.19 & 0.15 (0.55) & 0.34 & 10  \\ 
 p & 1.6 & 0.40 & 0.14 (0.33) & 0.54 & 35  \\
 p & 3.7 & 0.70 & 0.33 (0.88) & 1.03 & 21  \\
 p & 8.8 & 0.47 & 0.22 (0.80) & 0.69 & 10  \\
\end{tabular}
\end{ruledtabular}
\end{table}

If component separation were too extensive, our assumption about electron
wave functions would fail and the solid would become ordered over those
length scales which determine resistivity.  However, neutron star
crusts differ from white dwarf interiors in which component separation
is possible (see, for example Ref. \cite{schgim94}) because cooling to
$T \approx 0.5T_{m}$ occurs rapidly, in $10^{2-3}$s.  Vacancy mobility
(or its analogue in an amorphous system) is the dominant diffusion
mechanism enabling component separation to occur \cite{f72}.  Under
the assumption that sufficient sinks exist in the solid structure to
maintain a thermal equilibrium vacancy concentration during cooling,
the jump frequency for a given nucleus is of the order of
$\nu_{D}\exp(-\beta (H_{Fv} + \hat{\epsilon}))$, where $\nu_{D}$
is the Debye frequency and $\beta^{-1} = k_{B}T$.  In a {\it bcc}
lattice, the jump
activation energy given in Ref. \cite{f72} can be expressed,
approximately, as $\hat{\epsilon} \approx 15\mu V_{WS}\delta^{2}/4$,
in which $\mu$ is the shear modulus, $V_{WS}$ is the mean
volume per nucleus and $\delta^{2} = 0.08$.  Evaluation of this
expression, as an estimate of $\hat{\epsilon}$ for an amorphous system,
gives values much
smaller than calculated vacancy formation enthalpies $H_{Fv}$
\cite{j99} which are therefore the principal factor determining
diffusion rates.  For these formation enthalpies, jump frequencies
become negligible at $T < T_{m}$ relative to cooling rates.
A further consideration is the formation of clusters of like nuclei
which might form the initial stages of component separation.  Approximate
calculations \cite{j01} of the formation enthalpy of some specific
forms of cluster show that any decrease in enthalpy attained is
smaller than the
free energy term derived from configurational entropy at $T_{m}$.
On these grounds, we believe that component separation in neutron-star
matter is insignificant and that the amorphous nature of the solid
persists during cooling.  

The consequences of disorder resistivity are unambiguous.  From Eq. (1),
the exponential decay time for a mode of wavelength $\lambda$ is $\tau =
\sigma \lambda^{2}/\pi c^{2}$.  Under the assumption that $\sigma$ is
derived entirely from the high-density $\mathcal{R}$ of Table II, the
decay time is $\tau=\tau_{o}= 0.7-1.3 \times 10^{6}$ yr for a mode with
$\lambda/2 = 10^{5}$ cm, the approximate depth of the whole crust in a
$1.4 M_{\odot}$ neutron star.  This limits
the applicability of models in which the field is confined to the crust
\cite{pgz00} or has been submerged within it 
by post core-collapse accretion \cite{gpz99}.  Non-fossil field models,
in which considerable amplification occurs after formation of the neutron
star, rely on processes which are optimized at small electron density $n$.
This is true of thermoelectric amplification \cite{uy80,bah83,wg96} and
of amplification arising from the Hall term in Eq. (1) which requires
the inequality $\sigma B \gg nec$ \cite{gr92,rg02,hr04}.  Even if such
processes occur for the small $n$ at depths $\sim 10^{4}$ cm, the
models rely on movement of flux by Hall drift to regions of higher $n$
in which ohmic
dissipation is very slow.  Previous assumptions about $\sigma$ were
consistent with regions of this kind occupying most of the crust
volume.  But the magnitudes of resistivity produced by disorder give
ohmic dissipation times which are small everywhere in the spherical nuclear
phase of the crust \cite{pr95}.

For different limiting regions of $B$, solutions of Eq. (1) represent
either Hall drift or ohmic dissipation. A recent paper
\cite{caz04}, which also summarizes earlier work,
has defined the boundary between these regions as a function
of $Q$.  The conclusion was that, for $Q\stackrel{>}{\sim} 1$ and
$B\stackrel{<}{\sim} 10^{13}$ G, Hall drift never dominates the
evolution of currents in the crust.  Given these authors' assumed value
$\sigma \approx 4.4 \times 10^{25} Q^{-1}$ s$^{-1}$,
this is consistent with the
order of magnitude condition $\sigma B < nec$ being applied in the
high-density regions of the crust.
This Letter has shown that for amorphous structures, resistivity
is not defined directly by $Q$ but rather by the disorder which is an
inevitable consequence of $Z$-heterogeneity.  The specific values
of $\mathcal{R}$ calculated here allow one to reach, but with no free
parameter, the same conclusion as \cite{caz04}.  Thus field evolution
in neutron stars with $B < 10^{13}$ G, including almost the whole
population of radio pulsars, is governed by movement of flux from
the liquid core, if it occurs, and by ohmic dissipation in the crust.
The Maxwell stress is at least of the order of the critical shear stress
of the crust at depths $\stackrel{<}{\sim} 10^{4}$ cm.  Therefore
Eq. (1) may not be satisfied if, during the evolution of ${\bf B}$,
there are departures from hydromagnetic
equilibrium sufficiently large to cause mechanical failure of the solid.
But this complication does not affect the main conclusion of the present
Letter, which is that in neutron stars with ages exceeding a small
multiple of $\tau_{o}$, the magnetic flux density observed is a
fossil field, irrotational in the crust. 

The fields of $10^{14-15}$ G  thought to be present in the anomalous
X-ray pulsars (AXP) may evolve by Hall drift.  There are several possible
mechanisms for the persistent emission of these objects \cite{mcis02}
involving departures from hydromagnetic equilibrium in the star. 
But the resistivities of Table II mean that, unless the field is
almost irrotational
in the solid, ohmic dissipation must be a source of strong
thermal emission with a lifetime $\sim 10^{5-6}$ yr whether or not
hydromagnetic equilibrium exists.

\newpage 
\bibliography{apssamp}

\end{document}